\documentclass[preprint]{aastex}
\usepackage[utf8]{inputenc}

\usepackage[normalem]{ulem}
\usepackage{color}
\definecolor{red}{rgb}{1.0,0.0,0.0}

\shorttitle{GPI Observations of the AU Mic Debris Disk}
\shortauthors{Wang et al.}

\begin{document}
\slugcomment{accepted by ApJ Letters: August 18, 2015}
\title{Gemini Planet Imager Observations of the AU Microscopii Debris Disk: Asymmetries within One Arcsecond}

\author{
Jason J. Wang\altaffilmark{1}, 
James R. Graham\altaffilmark{1}, 
Laurent Pueyo\altaffilmark{2}, 
Eric L. Nielsen\altaffilmark{3,4}, 
Max Millar-Blanchaer\altaffilmark{5}, 
Robert J. De Rosa\altaffilmark{1}, 
Paul Kalas\altaffilmark{1}, 
S. Mark Ammons\altaffilmark{6}, 
Joanna Bulger\altaffilmark{7}, 
Andrew Cardwell\altaffilmark{8}, 
Christine Chen\altaffilmark{2}, 
Eugene Chiang\altaffilmark{1}, 
Jeffrey K. Chilcote\altaffilmark{9}, 
Ren\'{e} Doyon\altaffilmark{10}, 
Zachary H. Draper\altaffilmark{11,12}, 
Gaspard Duch\^{e}ne\altaffilmark{1,13}, 
Thomas M. Esposito\altaffilmark{14}, 
Michael P. Fitzgerald\altaffilmark{14}, 
Stephen J. Goodsell\altaffilmark{8,15}, 
Alexandra Z. Greenbaum\altaffilmark{16,2}, 
Markus Hartung\altaffilmark{8}, 
Pascale Hibon\altaffilmark{8}, 
Sasha Hinkley\altaffilmark{17}, 
Li-Wei Hung\altaffilmark{14}, 
Patrick Ingraham\altaffilmark{18}, 
James E. Larkin\altaffilmark{14}, 
Bruce Macintosh\altaffilmark{4,6}, 
Jerome Maire\altaffilmark{9}, 
Franck Marchis\altaffilmark{3}, 
Christian Marois\altaffilmark{12,11}, 
Brenda C. Matthews\altaffilmark{12,11}, 
Katie M. Morzinski\altaffilmark{19}, 
Rebecca Oppenheimer\altaffilmark{20}, 
Jenny Patience\altaffilmark{7}, 
Marshall D. Perrin\altaffilmark{2}, 
Abhijith Rajan\altaffilmark{7}, 
Fredrik T. Rantakyr\"{o}\altaffilmark{8}, 
Naru Sadakuni\altaffilmark{8}, 
Andrew Serio\altaffilmark{8}, 
Anand Sivaramakrishnan\altaffilmark{2}, 
R\'{e}mi Soummer\altaffilmark{2},
Sandrine Thomas\altaffilmark{18}, 
Kimberly Ward-Duong\altaffilmark{7}, 
Sloane J. Wiktorowicz\altaffilmark{21}, 
and Schuyler G. Wolff\altaffilmark{16,2}
}

\altaffiltext{1}{Astronomy Department, University of California, Berkeley, CA 94720}
\altaffiltext{2}{Space Telescope Science Institute, 3700 San Martin Drive, Baltimore MD 21218 USA}
\altaffiltext{3}{SETI Institute, Carl Sagan Center, 189 Bernardo Avenue, Mountain View, CA 94043, USA}
\altaffiltext{4}{Kavli Institute for Particle Astrophysics and Cosmology, Stanford University, Stanford, CA 94305, USA}
\altaffiltext{5}{Department of Astronomy \& Astrophysics, University of Toronto, Toronto ON M5S 3H4, Canada}

\altaffiltext{6}{Lawrence Livermore National Laboratory, 7000 East Ave., Livermore, CA 94040}
\altaffiltext{7}{School of Earth and Space Exploration, Arizona State University, PO Box 871404, Tempe, AZ 85287, USA}
\altaffiltext{8}{Gemini Observatory, Casilla 603, La Serena, Chile}
\altaffiltext{9}{Dunlap Institute for Astronomy \& Astrophysics, University of Toronto, 50 St. George St, Toronto ON M5S 3H4, Canada}
\altaffiltext{10}{Institut de Recherche sur les Exoplan\`{e}tes, D\'{e}partment de Physique, Universit\'{e} de Montr\'{e}al, Montr\'{e}al QC H3C 3J7, Canada}
\altaffiltext{11}{University of Victoria, 3800 Finnerty Rd, Victoria, BC, V8P 5C2, Canada}
\altaffiltext{12}{National Research Council of Canada Herzberg, 5071 West Saanich Road, Victoria, BC V9E 2E7, Canada}
\altaffiltext{13}{Université Grenoble Alpes / CNRS, Institut de Planétologie et d'Astrophysique de Grenoble, 38000 Grenoble, France}
\altaffiltext{14}{Department of Physics and Astronomy, UCLA, Los Angeles, CA 90095, USA}
\altaffiltext{15}{Center for Advanced Instrumentation, Durham University, Durham, DH1 3LE, UK}
\altaffiltext{16}{Physics and Astronomy Department, Johns Hopkins University, Baltimore MD, 21218, USA}
\altaffiltext{17}{University of Exeter, Astrophysics Group, Physics Building, Stocker Road, Exeter, EX4 4QL, UK}
\altaffiltext{18}{Large Synoptic Survey Telescope, 950N Cherry Av, Tucson AZ 85719, USA}
\altaffiltext{19}{Steward Observatory, 933 N. Cherry Ave., University of Arizona, Tucson, AZ 85721, USA}
\altaffiltext{20}{American Museum of Natural History, New York, NY 10024, USA}
\altaffiltext{21}{Department of Astronomy, UC Santa Cruz, 1156 High Street, Santa Cruz, CA 95064, USA}



\begin{abstract}
We present Gemini Planet Imager (GPI) observations of AU Microscopii, a young M dwarf with an edge-on, dusty debris disk. Integral field spectroscopy and broadband imaging polarimetry were obtained during the commissioning of GPI. In our broadband imaging polarimetry observations, we detect the disk only in total intensity and find asymmetries in the morphology of the disk between the southeast and northwest sides. The southeast side of the disk exhibits a bump at 1\arcsec\ (10~AU projected separation) that is three times more vertically extended and three times fainter in peak surface brightness than the northwest side at similar separations. This part of the disk is also vertically offset by 69$\pm$30~mas to the northeast at 1\arcsec\ when compared to the established disk mid-plane and consistent with prior ALMA and {\em Hubble Space Telescope}/STIS observations.
We see hints that the southeast bump might be a result of detecting a horizontal sliver feature above the main disk that could be the disk backside.
Alternatively when including the morphology of the northwest side, where the disk mid-plane is offset in the opposite direction $\sim$50~mas between 0\farcs4 and 1\farcs2, the asymmetries suggest a warp-like feature. Using our integral field spectroscopy data to search for planets, we are 50\% complete for $\sim$4~$M_\mathrm{Jup}$ planets at 4~AU. We detect a source, resolved only along the disk plane, that could either be a candidate planetary mass companion or a compact clump in the disk. 
\end{abstract}



\section{Introduction}
Debris disks are comprised of rocky and icy bodies, left over from planet formation.  The planetesimals in these disks collide, fragment, and produce micron-sized grains that scatter optical and near-infrared radiation from the star, resulting in a detectable nebulosity. Grains are influenced by the gravitational pull of larger bodies in the system as well as radiative forces and corpuscular drag due to the host star. These effects can induce morphology in debris disks such as midplane warps, as in $\beta$ Pictoris \citep{mouillet97a}, bow structures as seen in HD 61005 \citep{hines07}, and narrow, offset rings such as Fomalhaut \citep{kalas05a}.  

The debris disk around \object[AU Mic]{AU Microscopii} (AU Mic) is interesting for multiple reasons. At a distance of 9.9~pc, we are able to probe closer to the star than most systems. As a member of the $\beta$ Pictoris moving group \citep{barrado99}, its young age of 
$23\pm3$~Myr \citep{mamajek14}
makes it a prime candidate for directly detecting young planets still hot from their recent formation. It is also one of the very few M dwarfs with a debris disk imaged in scattered light. 
Since its ground-based discovery by \citet{kalas04}, this edge-on disk has been studied in scattered light using both adaptive optics and the \textit{Hubble Space Telescope} \citep{liu04,krist05,metchev05,graham07,fitz07,schneider14}, as well as in millimeter thermal emission with interferometric arrays \citep{wilner12,macgregor13}. From modeling these observations, it is believed AU Mic harbors a ``birth ring'' of planetesimals at $\sim$40~AU, inwards of which is empty and outwards of which is comprised of micron-sized dust particles placed on barely bound orbits \citep{strubbe06, augereau06}.
Within the disk, there are also 
local enhancements of surface brightness 
and deviations in the disk's vertical structure \citep[e.g.][]{fitz07, schneider14}. 
These features are likely non-axisymmetric structures located at or exterior to the birth ring and seen in projection \citep{graham07}.

The Gemini Planet Imager (GPI) is a high contrast imaging instrument on  Gemini South with a high-order adaptive optics (AO) system,
a coronagraph to suppress starlight, and 
a science instrument that features both integral field spectroscopy (spectral mode) and broadband imaging polarimetry (polarimetry mode) \citep{mac14,larkin14,perrin15_pol}. GPI gives us the opportunity to probe closer in and deeper than previous optical or near-infrared observations of AU Mic. We use GPI in both spectral and polarimetry modes to study the morphology of the debris disk and search for planets close in to the star.

\section{Observations and Data Reduction}
AU Mic was observed as part of commissioning the GPI instrument. On 2014 May 12, twenty-seven 60~s spectral mode frames were taken in the \textit{K1} spectral band (1.90--2.19~$\mu$m) at an airmass $<$ 1.01. Images of the sky and instrument thermal background were not taken with the observations, but thermal background data were taken earlier that night and used in the data reduction. Forty-four 60~s polarimetry mode observations were taken on 2014 May 15 in \textit{H}-band at an airmass $\leq$ 1.015. The GPI waveplate was rotated by $22\fdg5$ between frames. AU~Mic transits almost directly overhead at Cerro Pachon, so a total of $154^{\circ}$ of field rotation was obtained in each data set for Angular Differential Imaging \citep[ADI,][]{marois06}. For the \textit{H}-band data, four frames were discarded due to anomalies associated with  commissioning tests. 

The raw data consist of individual spectra in spectral mode and pairs of spots for the two orthogonal polarizations dispersed by the Wollaston prism in polarimetry mode. Each spectrum or spot pair corresponds to a spatial resolution element, or spaxel, in a datacube created by the GPI Data Reduction Pipeline \citep[DRP,][]{perrin14_drp}. The spectral mode data were dark subtracted, corrected for shifts in the position of individual spectra on the detector due to instrument flexure \citep{wolff14} with an \textit{H}-band argon arc lamp exposure taken immediately before the observing sequence, wavelength calibrated with \textit{K1}-band argon arc lamp data, fixed for bad pixels in the 2-D data, assembled into a spectral data cube, thermal background subtracted, fixed for bad pixels in the 3-D datacube, and corrected for distortion. 

The reduction of the polarimetry mode data follows the steps described in \citet{perrin15_pol}. To create polarimetry datacubes where the third dimension contains two orthogonal polarizations, the data were dark subtracted, flexure corrected using a cross-correlation routine \citep{draper14}, fixed for bad pixels in the 2-D data, assembled into a datacube using a model of the point spread function of the spot pair, fixed for bad pixels in the 3-D datacube, and corrected for distortion. The polarimetry datacubes were then collapsed in the polarization dimension to obtain 2-D total intensity images. For our total intensity reductions, we do not construct the full Stokes datacube, but one is made with the GPI DRP to search for a polarized signal.

To combine and photometrically calibrate our data, we used the GPI DRP to measure the four fiducial diffraction or ``satellite'' spots, which are centered on the occulted star and imprinted with its attenuated spectrum \citep{wang14}. In spectral mode, the satellite spot fluxes were used to calibrate the data photometrically relative to the occulted star. For each datacube, the location of the occulted star at each wavelength was found using a least squares fit to all of the satellite spots' positions and the magnitude of the atmospheric differential refraction. In polarimetry mode, the satellite spot flux calibration is an ongoing task, so no flux-calibrated polarimetry mode data is presented here. The central star in the polarimetry data was located using a radon-transform-based algorithm \citep{wang14, pueyo15}. The precision on the star center is 0.7~mas and 0.9~mas for spectral and polarimetry mode respectively \citep{wang14}. 

To remove the stellar point spread function (PSF), we used \texttt{pyKLIP} \citep{wang15}, a Python implementation of the Karhunen-Lo\`eve Image Projection (KLIP) algorithm \citep{soummer12,pueyo15}. To distinguish astrophysical sources from the stellar PSF, we used ADI for the polarimetry mode data and both ADI and spectral differential imaging \citep[SDI,][]{marois00} for the spectral mode data. Multiple PSF-subtracted images were made depending on the purpose of the reduction. For all the subtractions, we divided the PSF into annuli, divided each annulus into azimuthal sectors, and ran KLIP on each sector. 

The main parameters we adjusted were the number of modes used from the Karhunen-Lo\`eve (KL) transform and an exclusion criteria for reference PSFs, similar to $N_{\delta}$ in \citet{laf07}, defined by the number of pixels a hypothetical astrophysical source would move azimuthally, radially or both due to ADI/SDI. For the \textit{H}-band total intensity data, we used three KL modes and a ten-pixel exclusion for the diffuse left side of the disk presented in Figure \ref{fig:fourdisks}(a) and four KL modes and a six-pixel exclusion for the more compact right side presented in Figure \ref{fig:fourdisks}(a). We also searched for fainter features using ten KL modes and a three-pixel exclusion as shown in Figure \ref{fig:fourdisks}(b). For the \textit{K1}-band data, we used ten KL modes and a three pixel exclusion criteria to characterize a compact candidate source shown in Figure \ref{fig:fourdisks}(c) and twenty KL modes and a 1.5 pixel exclusion criteria to filter out extended disk features and quantify our sensitivity to point sources in Section \ref{sec:planet}.

The frames were rotated to orient the disk horizontally using a disk position angle (PA) of $128\fdg41$ \citep{macgregor13} derived from observations with Atacama Large Millimeter/submillimeter Array (ALMA) before being mean combined to produce the images in Figure \ref{fig:fourdisks}.

We did not detect any polarized signal (5\% upper limit on the polarization fraction at 0\farcs5), consistent with a low expected polarization for forward scattered light at small projected angular separations \citep{graham07} and reduced polarization sensitivity due to image persistence in the HAWAII 2-RG detector \citep{smith08}, for which calibration work in ongoing.

\section{Analysis of the Debris Disk \label{sec:disk}}

All three reductions in Figure \ref{fig:fourdisks} show stark asymmetries between the southeast (SE) and northwest (NW) sides of the disk. To quantify these asymmetries, we used the conservative \textit{H}-band reduction in Figure \ref{fig:fourdisks}(a) and fit Gaussian vertical profiles to the disk at various separations. We fit between 0\farcs4 and 1\farcs25. The inner radius is determined by the increasing noise at small angular separations and the outer radius is set by GPI's finite field of view. Due to the relatively low signal-to-noise and ADI self-subtraction in the vertical wings of the disk, a Gaussian provides good fits with well-behaved residuals and performs as well as alternate profiles such as the Lorentzian used by \citet{graham07}. 
The primary aim of the Gaussian fit was to measure the disk’s mid-plane vertical offset and full-width-half-maximum (FWHM). We also obtained the peak brightness of the disk as part of the fit and calculated the vertically integrated flux by summing over the flux within one FWHM centered about the measured disk mid-plane.

Our reductions are subject to varying levels of disk self-subtraction, which biases our measurements. To correct for this effect, we repeated the analysis with model disks injected separately at one of five non-overlapping positions, resulting in ten sides for analysis. The disk model is a symmetric, edge-on, optically-thin ring parameterized by a power-law grain size distribution, a Henyey-Greenstein scattering function, and a Gaussian vertical profile. We adopted values from the porous water-ice model from \citet{graham07} and varied both the scale height and the surface brightness to accommodate the range of measured values. As GPI only sees the inner arcsecond of the disk, the main contribution to the flux in the models is the front edge of the disk with scattering angles $< 15^{\circ}$. The mean FWHM, peak brightness and integrated vertical flux we measured from the model disks were used to correct biases in our measured values as a function of separation. The scatter of all four quantities in the model disks were used as the uncertainties in our measurements.

\begin{figure}

\centering
\includegraphics[trim = 20mm 5mm 20mm 5mm, clip, scale=0.60]{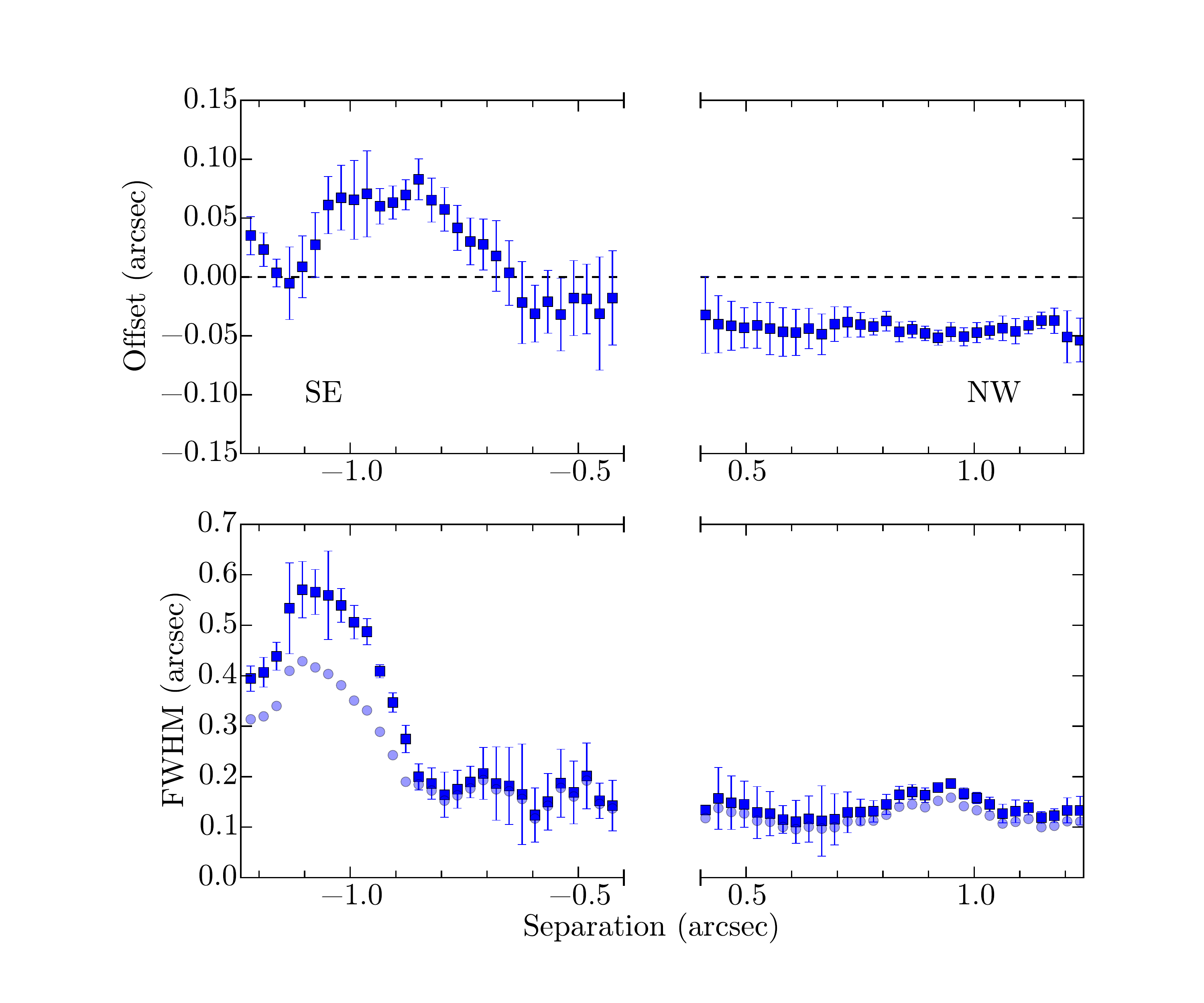}
\caption{Measurements of the morphology of the disk in \textit{H}-band with the negative and positive separations for the SE and NW sides respectively. At each separation, we plot the disk mid-plane vertical offset and FWHM with 1-$\sigma$ error bars as blue squares. The faded circles represent the FWHM before correcting for biases. Adjacent points are correlated so that the plot is well sampled. 
\label{fig:fourplots}}
\end{figure}

We present measurements of the mid-plane offset and FWHM as a function of separation with 1-$\sigma$ error bars in Figure \ref{fig:fourplots}. 
The vertical mid-plane offset peaks around 1\arcsec\ on the SE side with an offset from the average ALMA mid-plane of 69$\pm$30~mas to the northeast at 1\arcsec\ separation. 
The mid-plane offset is roughly constant between 0\farcs4 and 1\farcs2 on the NW side, with an offset of 47$\pm$8~mas to the southwest at 1\arcsec. The FWHM is roughly constant on the NW side and consistent with measurements at larger separations but triples on the SE side past 0\farcs8 to a peak FWHM of 570$\pm$56~mas. Due to unsubtracted residual PSF features, the disk brightness as a function of separation that we measure is not reliable. However, the overall brightness asymmetry between the two sides as seen in Figure \ref{fig:fourdisks} is real: on average, the peak surface brightness is three times larger on the NW side, but the vertically integrated brightnesses are consistent with being equal.

From the aggressive reduction in Figure \ref{fig:fourdisks}(b), we see a dark gap and a faint horizontal sliver on the SE side, above the main disk. This sliver is only at 3.1$\sigma$ significance when comparing the total flux of this feature to randomly summing up noise in the same manner. However, it emerges in successive reductions that progressively remove speckle noise and does not point radially toward the star, like residual speckles. We cannot rule out that the apparent morphology is influenced by KLIP self-subtraction. We also do not observe any indication of a corresponding feature on the NW side, which could be hidden underneath the over-subtraction of the brighter main belt of the disk. 

\section{Sensitivity to Planets\label{sec:planet}}

\begin{figure}
\epsscale{.75}
\plotone{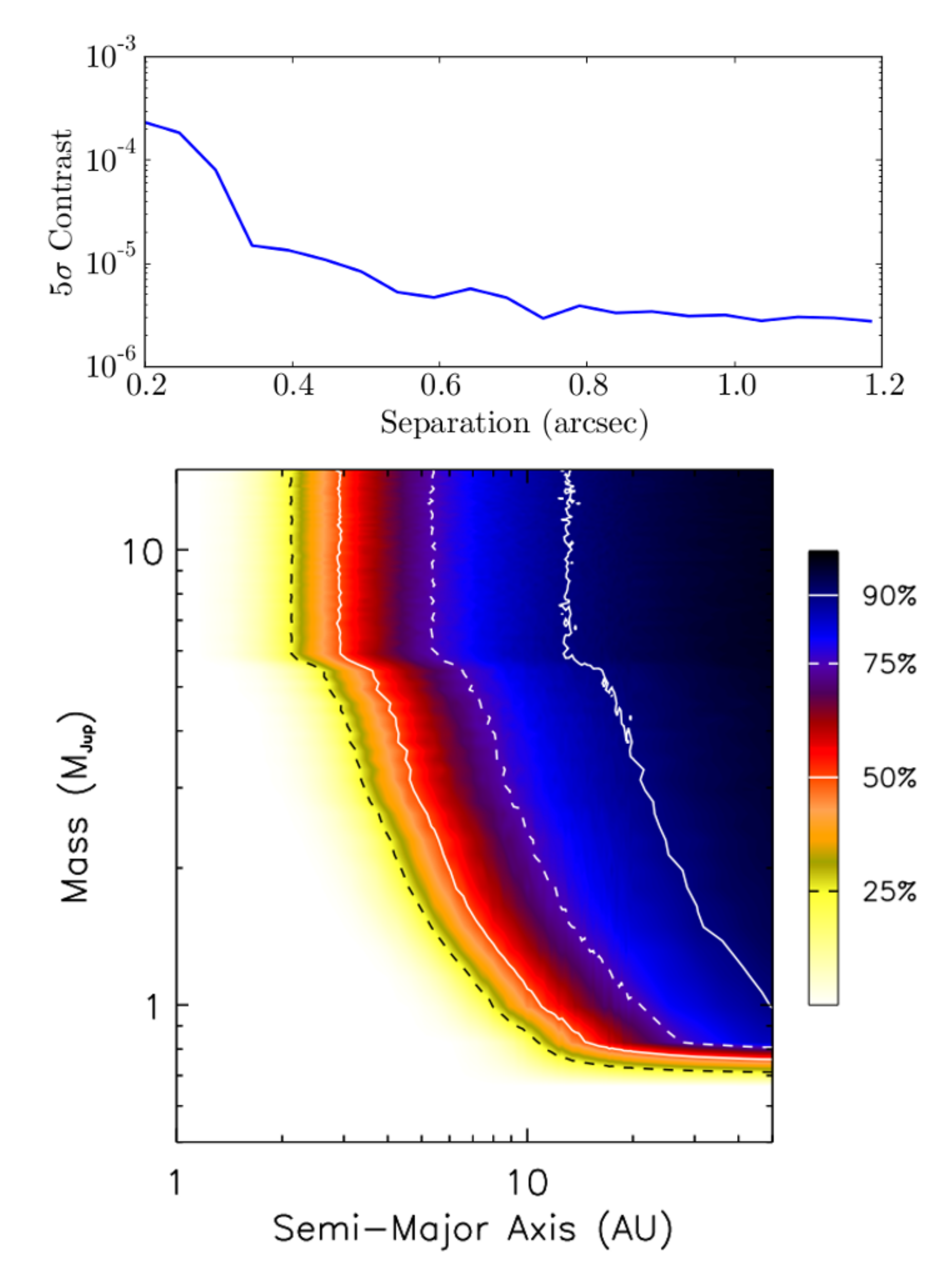}
\caption{\textit{(Top)} Sensitivity to point sources as a function of projected separation using the \textit{K1}-band spectral mode data. \textit{(Bottom)} Completeness for planets as a function of semi-major axis and mass in the AU Mic system based on contrast curves from NICI \citep{wahhaj13} and GPI (this paper). Completeness is calculated using a Monte Carlo method described by \citet{nielsen} where planets with random orbits are generated, the contrast curves determine whether they are detected, and COND models \citep{baraffe03} are used to convert from luminosity to mass. \label{fig:completeness}}
\end{figure}

The achieved sensitivity for our \textit{K1} data was calculated following the procedure outlined by \citet{mawet14} and the $5\sigma$-equivalent false positive threshold is plotted in Figure \ref{fig:completeness}. Flux attenuation due to PSF subtraction was quantified by injecting and recovering the brightnesses of simulated planets. To translate our contrast curve to limits on possible planets in the system, we ran a Monte-Carlo analysis as described by \citet{nielsen} to determine the completeness of our data. Since GPI is not sensitive to planets at large orbital distances due to its restricted field of view, we also included the limits from the Near-Infrared Coronagraphic Imager \citep[NICI,][]{wahhaj13} in Figure \ref{fig:completeness}. With the improved inner working angle, the GPI data are 50\% complete for 4~$M_{\rm Jup}$ planets with semi-major axes of 4~AU assuming COND models \citep{baraffe03}.

We did find one compact \textit{K1} candidate, marked in Figure \ref{fig:fourdisks}(c), at a separation of 544$\pm$4~mas, a PA of $130\fdg2$$\pm$$0\fdg9$, and a \textit{K1} contrast of 1.1$\pm$0.3~$\times 10^{-5}$. Although this source is detected with 5-$\sigma$ confidence in \textit{K1}, there is no corresponding source at \textit{H}. Because polarimetry mode flux calibration is still ongoing, we cannot currently quantify the constraints placed by our \textit{H}-band data. 
According to COND models \citep{baraffe03}, this source would  correspond to a 2-3 $M_{\rm Jup}$ planet. However, inspection of Figure \ref{fig:fourdisks}(c) reveals that the clump is extended along the disk. The FWHM perpendicular to the disk is $\simeq$70~mas and consistent with being unresolved when compared to the satellite spot PSF; parallel to the disk the measured FWHM is $\simeq$160~mas giving an intrinsic size of $\simeq$140~mas (1.4~AU), suggestive of a local enhancement in the disk similar to those reported at larger projected separations \citep[e.g.,][]{liu04,fitz07,schneider14}. We are confident in the measured elongation because it is significantly larger than a residual speckle, which, given the \textit{K1} bandwidth, would subtend about 100~mas at this angular separation.

\section{Discussion}

The AU Mic debris disk has significant asymmetries in the morphology and surface brightness between the SE and NW sides of the disk. In Figure \ref{fig:alma}, we compare our observations with 1.3-mm continuum data from ALMA \citep{macgregor13} and optical imaging using the using the Space Telescope Imaging Spectrograph (STIS) on the {\it Hubble Space Telescope} \citep{schneider14}. The ALMA data probe larger planetesimals insensitive to the variable radiation pressure and stellar wind drag of AU Mic, making it easier to disentangle dynamical and radiative forces on the debris disk. STIS, like GPI, traces scattered light from small grains, but with a larger field of view. 

The disk mid-plane offsets, FWHM enhancement, and brightness asymmetries we find agree well with the published STIS results from \citet{schneider14}. The most prominent common feature is the faint, extended, and vertically-offset bump on the SE side at 1\arcsec. This feature is also continuous in the STIS data where it is seen to taper back down near 2\arcsec\ from the star. Since the SE bump is also consistent with the most prominent undulation in the ALMA contours, it is plausible that the feature is associated with gravitational perturbations of the parent bodies. The statistical significance of this mm feature is questioned by \citet{macgregor13}; however, the correspondence between ALMA and shorter wavelength observations lends credence to its reality.  Assuming the mm grains are not displaced from the birth ring, the correlation between scattered light and mm-emission suggests that the SE bump is located in the $\sim$40~AU birth ring and is just seen projected at $\sim$10~AU separation. As both the STIS and ALMA data were taken within four years of the GPI observations, we cannot yet distinguish between orbital motions at 10 and 40~AU. 

The compact \textit{K1} candidate in our data could be another clump of disk material or a planet with a circumplanetary disk. The candidate must be a transient feature unless the object is self-gravitating. If the projected size of this source is representative of its radial extent ($\delta a$), Keplerian shear at an orbital separation ($a$) of 40~AU will completely erase this feature in only $(2/3) (a/\delta a) = 19 (a/40~{\rm AU}) (1.4~{\rm AU}/\delta a)$ orbits. However, a 1.4~AU Hill sphere at 40~AU is established by a $0.08 (M_*/0.6 M_\odot)~M_\mathrm{Jup}$ object, which is well below our detection limits. Follow up observations are required though for verification and to determine common proper motion.

\begin{figure}
\plotone{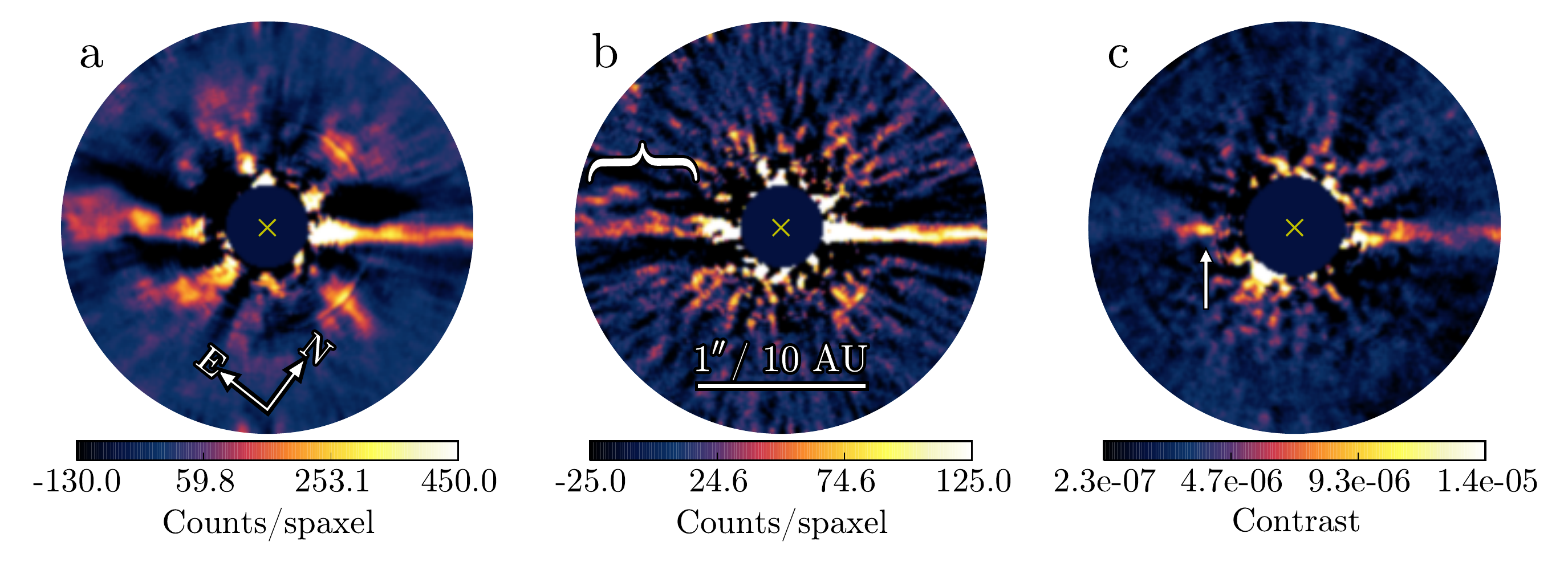}
\caption{Reduced images of the AU Mic disk. The images are rotated so that the disk is horizontal. Images (a) \& (b) are from the \textit{H}-band total intensity broadband observations. Image (a) is optimized for the bright main belt of the disk. The azimuthal structure outside of the horizontal disk plane is due to unsubtracted speckles. Image (b) is used to look for fainter features. The curly brace indicates the position of the sliver feature we detect. Image (c) is from the \textit{K1}-band spectral mode observations. The arrow indicates the candidate compact source that we detect. \label{fig:fourdisks}}

\end{figure}

\begin{figure}
\epsscale{1.0}
\plotone{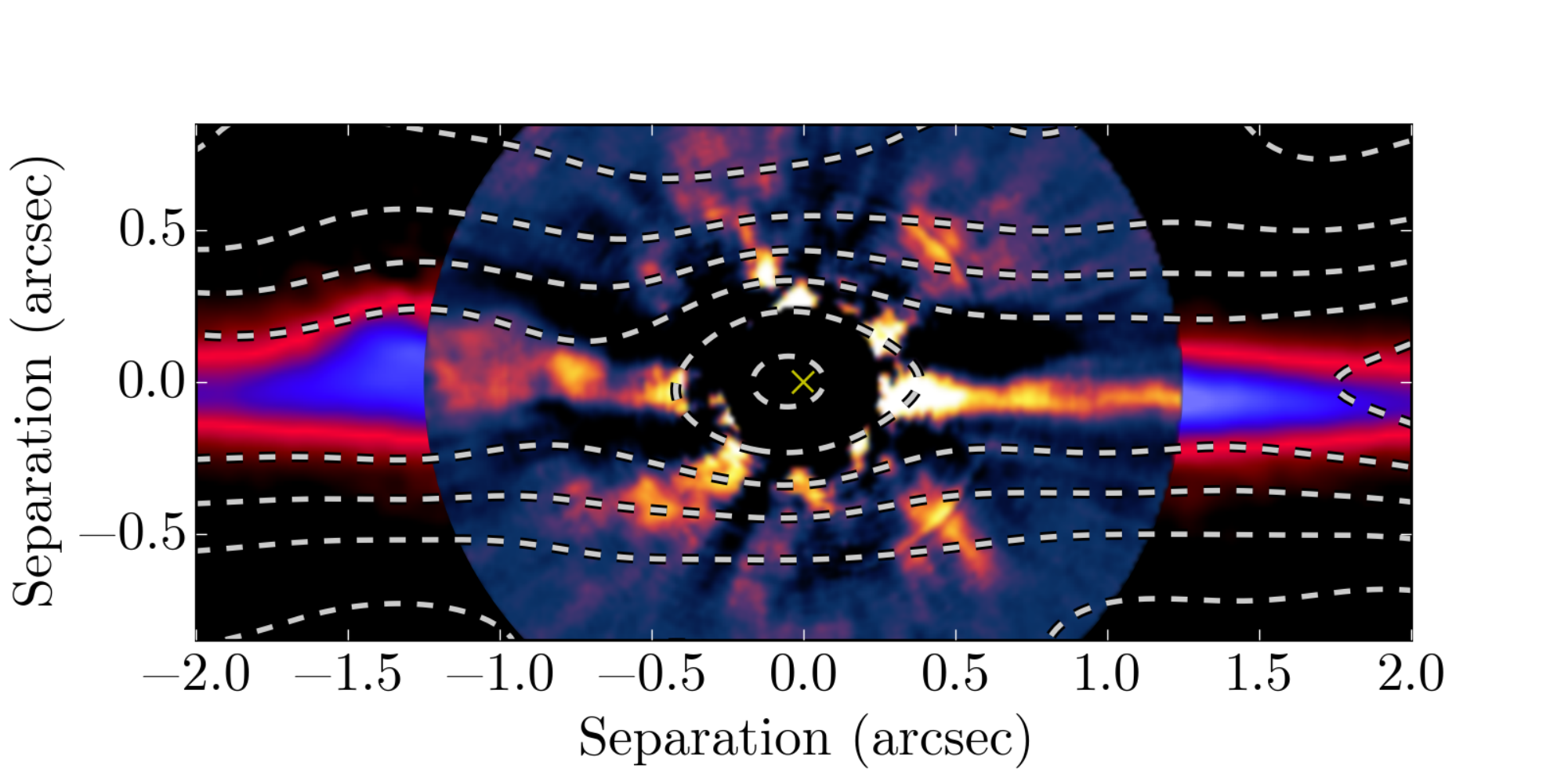}
\caption{Comparison of GPI data with previous observations. GPI \textit{H}-band data from Figure~\ref{fig:fourdisks}(a) is shown in the center with contours from 1.3-mm ALMA observations \citep{macgregor13} overlaid on top using a linear scaling and an image from STIS \citep{schneider14} in the background. The southeast (left) bump between $-1.5$\arcsec\ and $-1$\arcsec\ in the disk is apparent in all three observations. The mid-plane offset is also continuous between the STIS and GPI data. \label{fig:alma}}
\end{figure}

Ignoring the possible detection of the sliver in Figure \ref{fig:fourdisks}(b), the vertical mid-plane offset variations on the two sides of the disk in opposite directions hint at a vertical warp in the disk. Applying the analytical model from \citet{dawson11} to the AU Mic system, any inclined planet between an Earth and Neptune mass could induce the warp depending on the orbital separation and dynamical timescale. The candidate companion we see at a projected separation of 5.5~AU can be slightly lower than an Earth mass and still be responsible for the warp. The FWHM and flux asymmetries in the disk could be due to dynamical heating from planets or parent bodies in the ring, which can puff up the disk vertically but keep the total amount of starlight scattered constant. 

If the sliver feature is real, it may be the backside of the disk (the bright main belt would be the front side). In this case, the variations in the disk could be explained by an inclined ring with azimuthally varying brightness due to clumps in the disk, like those seen at further separations \citep{fitz07}. Essentially, on the SE side, the backside is visible due to a local brightness enhancement that is absent on the NW side. However, without a corresponding detection on the northwest, the backside interpretation is uncertain.

Our GPI observations show that there remain many unanswered questions regarding this extensively studied debris disk. The stark asymmetries between the two sides of the disk and correlation with millimeter observations hint at gravitational interactions due to perturbers. With deeper imaging by new near-infrared high contrast imaging instruments and higher resolution data in the millimeter by ALMA, we may soon have a much better understanding of the physics in the system. 

\acknowledgments
{\bf Acknowledgements:}  We thank M. MacGregor and G. Schneider for making available to us the reduced ALMA and STIS data respectively. 
This research was supported in part by NASA NNX15AD95G, NASA NNX11AD21G, NSF AST-0909188, NSF AST-1413718, and the University of California LFRP-118057.
The GPI project has been supported by Gemini Observatory, which is operated by AURA, Inc., under a cooperative agreement with the NSF on behalf of the Gemini partnership: the NSF (USA), the National Research Council (Canada), CONICYT (Chile), the Australian Research Council (Australia), MCTI (Brazil) and MINCYT (Argentina). This research has made use of the SIMBAD database, operated at CDS, Strasbourg, France.

{\it Facilities:} \facility{Gemini South (GPI)}.



\end{document}